\documentclass[9pt,twocolumn,twoside]{optica}
\usepackage{ulem}
\usepackage{soul}
\setboolean{shortarticle}{false}
\setboolean{minireview}{false}

\title{Periodically poled thin film lithium niobate microring resonators with a second-harmonic generation efficiency of 250,000\,\%/W }

\author[1]{Juanjuan Lu}
\author[1]{Joshua B. Surya}
\author[1]{Xianwen Liu}
\author[1]{Alexander W. Bruch}
\author[1]{Zheng Gong}
\author[1]{Yuntao Xu}
\author[1,*]{Hong X. Tang}

\affil[1]{Department of Electrical Engineering, Yale University, New Haven, Connecticut 06511, USA}

\affil[*]{Corresponding author: hong.tang@yale.edu}

\begin{abstract}
Lithium niobate (LN), dubbed by many as the silicon of photonics, has recently risen to the forefront of chip-scale nonlinear optics research since its demonstration as an ultralow-loss integrated photonics platform. Due to its significant quadratic nonlinearity ($\chi^{(2)}$), LN inspires many important applications such as second-harmonic generation (SHG), spontaneous parametric down-conversion, and optical parametric oscillation. Here, we demonstrate high-efficiency SHG in dual-resonant, periodically poled z-cut LN microrings, where quasi-phase matching is realized by field-assisted domain engineering. Meanwhile, dual-band operation is accessed by optimizing the coupling conditions in fundamental and second-harmonic bands via a single pulley waveguide. As a result, when pumping a periodically poled LN microring in the low power regime at around 1617\,nm, an on-chip SHG efficiency of 250,000\,\%/W is achieved, a state-of-the-art value reported among current integrated photonics platforms. An absolute conversion efficiency of 15\% is recorded with a low pump power of 115\,$\mu$W in the waveguide. Such periodically poled LN microrings also present a versatile platform for other cavity-enhanced quasi-phase matched $\chi^{(2)}$ nonlinear optical processes.
\end{abstract}

\setboolean{displaycopyright}{true}

\begin{document}

\maketitle

\section{Introduction}
Second-order nonlinearity ($\chi^{(2)}$) is the basis of many important nonlinear optical processes such as second-harmonic generation (SHG), sum and difference frequency generation and parametric down-conversion. Among these, SHG is specifically important for a range of applications including precision frequency metrology \cite{RevModPhys.75.325,synthesizer}, optical clocks \cite{Newman:19}, molecular imaging \cite{SHG:image, Han:05}, and quantum-information processing \cite{xiang:light, quantuminfo}. Compared with traditional SHG in bulk materials \cite{161322}, the advanced nanophotonic techniques have enabled SHG in a miniaturized and power-efficient microchip based on a variety of integrated photonics platforms, such as silicon nitride \cite{Levy:11, xxxue:17}, gallium arsenide \cite{doi:10.1002/lpor.201800149,kuo2016}, aluminum nitride (AlN) \cite{Guo:16, Surya:18, alexSHG}, lithium niobate (LN) \cite{chang2016thin,Wolf:18,PhysRevApplied.11.034026,Luo:18, Wang:18,doi:10.1002/lpor.201800288},  and others \cite{Xiong:11,Roland:16,Scaccabarozzi:06,Rivoire:09,Yamada:14}. Among these, LN is particularly attractive due to its large $\chi^{(2)}$ nonlinearity, a broad transparency window from 350\,nm to 4.5\,$\mu$m, flexibility in ferroelectric domain control as well as recent advances in the development of low-loss thin-film LN on insulator (LNOI) platform \cite{chang2016thin,Wolf:18,Luo:18, Wang:18,doi:10.1002/lpor.201800288,Zhang17,wang18,zhang19,wang19,eocomb,he2018self,Desiatov:19,Luo:17,PhysRevApplied.11.034026,Yu:19,Lu:19}. 

Leveraging the small mode volume and large power enhancement in high quality-factor (Q) optical microcavities \cite{vahalamicrocavity}, high-efficiency SHG with low power consumption can be envisioned. In recent decades, cavity-enhanced SHG in LN integrated photonics platform has been reported through the techniques of modal (MPM) \cite{PhysRevApplied.11.034026,Wang:14,Chen:18}, cyclic (CPM) \cite{PhysRevApplied.6.014002,Luo:17,PhysRevLett.122.173903}, and quasi-phase matching (QPM) \cite{Wolf:18}. The latter predicts a higher SHG efficiency as it involves two fundamental modes for a larger modal overlap. Meanwhile, QPM is generally achieved through the periodic poling and hence allows the engineering of the phase-matching wavelengths for specific applications. The potential of the LNOI platform, however, is far from being realized as the demonstrated cavity-enhanced-SHG efficiencies of recent works remain much lower than the theoretical value \cite{PhysRevApplied.11.034026} as well as those achieved in integrated AlN platforms with a lower $\chi^2$ coefficient \cite{alexSHG}. 

In this paper, we investigate QPM-based SHG in waveguide-coupled, periodically poled lithium niobate (PPLN) microring resonators. Through the design and fabrication optimization, the PPLN microring exhibits high intrinsic Q-factors, and efficient dual-band coupling via a single pulley waveguide. Intense electric fields are applied to invert the crystal domain of z-cut LN microrings periodically with high fidelity and thereby enable first-order QPM. Additionally, the detuning between the pump and second-harmonic (SH) cavity modes is compensated by tuning the chip temperature. As a result, from a fully integrated PPLN microring device, we are able to achieve an on-chip SHG efficiency of up to 250,000\,\%/W in the low pump power regime. An absolute power conversion efficiency as high as 15\% is also attained with 115\,$\mu$W pump power in the waveguide.  

\section{Principles of Device Design and Fabrication}
\label{sec:design}
\begin{figure}[htbp]
\centering
\includegraphics[width=\linewidth]{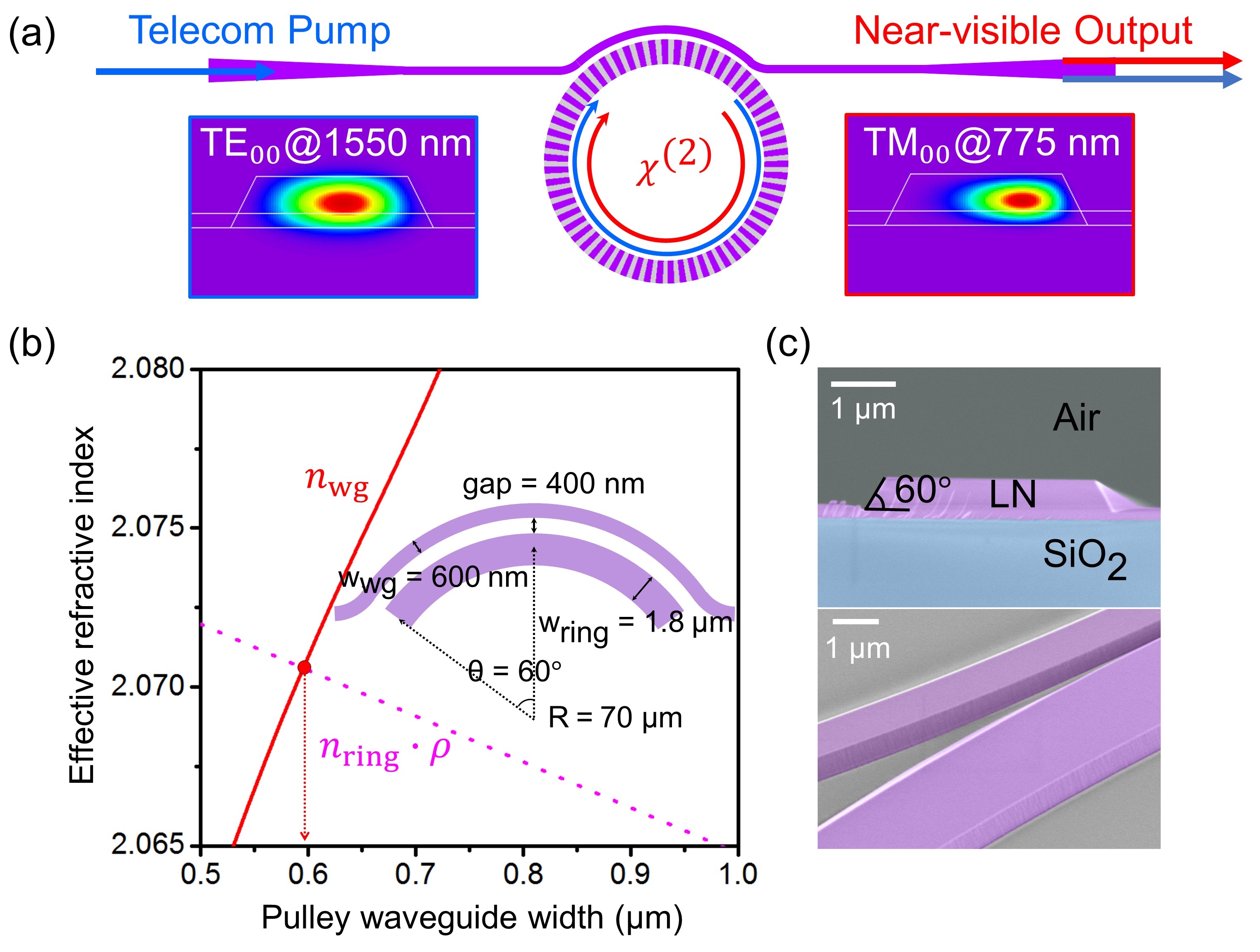}
\caption{(a) A schematic of the PPLN microring resonator. The pump telecom TE$_{00}$ and SH near-visible TM$_{00}$ modes in the microring are excited through a pulley coupling waveguide. The corresponding simulated mode profiles are shown. (b) Simulated effective refractive indices of the TM$_{00}$ mode at 775\,nm for the LN microring (dashed pink line) and pulley waveguide (solid red line) by a finite difference method (FDM). The inset details the coupling geometry. (c) False-color SEM images of the cleaved waveguide facet (top) and the sidewall around the waveguide-microring coupling region (bottom).}
\label{fig:design}
\end{figure}

Figure\,\ref{fig:design}(a) depicts the design principle of the PPLN microring, where the SHG process involves the fundamental transverse electric (TE$_{00}$) and magnetic (TM$_{00}$) optical modes of angular frequencies $\omega_{a}$ and $\omega_{b}$ in the telecom and near-visible bands, respectively. Pumping at an angular frequency $\omega_\mathrm{p}$, the on-chip SHG efficiency in the non-depletion regime \cite{Guo:16} is given by
\begin{equation}
\eta=\frac{P_{SHG}}{P_{p}^{2}}=\frac{g^{2}\omega_{b}}{\hbar \omega_{p}^2} \frac{2\kappa_{c, b}}{\delta_{b}^2+\kappa_{b}^{2}} \left(\frac{2\kappa_{c, a}}{\delta_{a}^2+\kappa_{a}^{2}}\right)^2,
\label{eq1}
\end{equation}
where $P_{p}$($P_{SHG}$) denotes the on-chip pump (SHG) power; $\kappa_{c}$, $\kappa_0$ and $\kappa$ are respectively the external, intrinsic and total coupling rates of the cavity modes with $\kappa =\kappa_{c}+\kappa_0$. Meanwhile, $\delta_{a}=\omega_{a}-\omega_{p}$ ($\delta_{b}=\omega_{b}-2\omega_{p}$) is the detuning for mode $a(b)$. The nonlinear coupling strength $g$ between mode $a$ and $b$ reads \cite{Guo:16},
\begin{equation}
g \approx \sqrt{\frac{\hbar \omega_{a}^{2}\omega_{b}}{\varepsilon_{0} 2\pi R}}\cdot\frac{\zeta}{\varepsilon_{a}\,\sqrt{\varepsilon_{b}}}\cdot \frac{3\chi^{(2)}_\mathrm{eff,N}}{4\sqrt{2}} \cdot\delta\left(m_{b}-2 m_{a}-M\right).
\label{eq2}
\end{equation}
Here, $\varepsilon_0$ is the vacuum permittivity, $R$ is the microring radius, $\varepsilon_{a(b)}$ and $m_{a(b)}$ are the effective relative permittivity and azimuthal number of mode $a(b)$ in the microring, and the integer $M=\frac{2 \pi R}{\Lambda}$ is defined by the poling structure with $\Lambda$ being the poling period. When the QPM condition $m_{b}-2 m_{a}-M = 0$ is satisfied, non-zero coupling strength $g$ can then be accessed. The effective quadratic tensor element $d_\mathrm{eff}$ of the LN material and the poling duty cycle $D$ determine the effective $\chi^{(2)}$ susceptibility of the PPLN microring with $N$-th order QPM, as denoted by $\chi_\mathrm{eff,N}^{(2)}$, through the relation \cite{HUM2007180,Wolf:18}
 \begin{equation}
 \chi_\mathrm{eff,N}^{(2)}=2 \cdot d_\mathrm{eff,N}=4 \cdot d_\mathrm{eff} \cdot \frac{sin(\pi N D)}{N \pi}.
 \label{eq3}
 \end{equation}
The modal overlap factor $\zeta$ is expressed as
\begin{equation}
\zeta=\frac{\iint\mathrm{d}r\mathrm{d}z\left(u_{a,r}^{*}(r,z)\right)^{2} u_{b,z}(r,z)}{\iint\mathrm{d}r\mathrm{d}z\, |u_{a,r}(r,z)|^2\cdot \sqrt{\iint\mathrm{d}r\mathrm{d}z\, |u_{b,z}(r,z)|^2}}, 
\label{eq4}
\end{equation} 
where $u_{a(b),r(z)}(r,z)$ is the corresponding electric field component of cavity mode $a(b)$. 

According to the analysis above, when the critical coupling ($\kappa_{c,a(b)}$\,=\,$\kappa_{0,a(b)}$) and dual-resonance ($\omega_{p}$\,=\,$\omega_{a}$\,=\,$\frac{\omega_{b}}{2}$) conditions for both pump and SH modes are fulfilled, a maximum SHG conversion efficiency can be derived as \begin{equation}
\eta_{\mathrm{max}}=\frac{g^{2} Q_{0,a}^{2} Q_{0,b}}{\hbar \omega_{a}^{4}},
\label{eq5}
\end{equation}
where $Q_{0,a(b)}$ is the intrinsic Q-factor of mode $a(b)$ and is determined by $Q_{0,a(b)}=\frac{\omega_{a(b)}}{2 \cdot \kappa_{0,a(b)}}$. It is evident that high intrinsic Q-factors and large nonlinear coupling strength $g$ are critical for achieving $\eta_{\mathrm{max}}$ in a QPM-based microring. In the following subsections, we outline two core design procedures: (1) realization of dual-band critical coupling and high-$Q_0$ LN microring; (2) implementation of periodic poling yielding a large $g$ while satisfying the QPM and dual-resonance conditions.
\subsection{Coupling design and high-Q LN microring fabrication}
The wide spectral separation between the telecom pump and near-visible SH modes poses a challenge for efficient dual-band waveguide-microring coupling using a conventional straight waveguide (point) coupler due to the significantly weaker coupling strength at shorter SH wavelengths. Compared with the MPM case with an extra SH extraction waveguide \cite{Guo:16,Surya:18,alexSHG}, QPM-based SHG involves fundamental telecom and near-visible modes and allows us to employ a single pulley waveguide shown in Fig.\,\ref{fig:design}(a) to address that issue. For a pulley bus waveguide, the coupling strength depends on both the cross-sectional evanescent field overlap and the effective refractive index matching between the waveguide and cavity modes given by $n_\mathrm{wg}=n_\mathrm{ring}\,\cdot\,\rho$ \cite{Hosseini:10}. The coefficient $\rho\,=\frac{R+w_\mathrm{ring}/4}{R+\mathrm{gap}+(w_\mathrm{wg}+w_\mathrm{ring})/2}$ denotes the ratio between the two path lengths, with $w_\mathrm{wg}$ and $w_\mathrm{ring}$ being the width of the waveguide and microring, respectively. Considering the larger evanescent field overlap of the telecom modes, we specifically optimize $w_\mathrm{wg}$ to satisfy the aforementioned index matching condition in the near-visible band and thereby access efficient dual-band coupling. 

As shown in Fig.\,\ref{fig:design}(b), the optimal $w_\mathrm{wg}$ is simulated to be approximately $600\,\mathrm{nm}$ for the fixed $w_\mathrm{ring}$ of $1.8\,\mu\mathrm{m}$ and $R$ of $70\, \mu \mathrm{m}$. Meanwhile, the thickness of the LN waveguide is $600\,\mathrm{nm}$ with an unetched bottom layer of $180\,\mathrm{nm}$ and the coupling gap is set to be $400\,\mathrm{nm}$. The large ring width and shallow etched geometry are chosen to mitigate the scattering losses from the sidewall roughness \cite{Liu:17}. The LN microrings were fabricated on a z-cut undoped congruent LNOI wafer (from NANOLN), and the device parameters (e.g., $w_\mathrm{wg}$, gap, and $\theta$) were experimentally optimized to enable the dual-band critical coupling. The fabrication process is presented in a previous work \cite{Lu:19}. Figure\,\ref{fig:design}(c) shows the false-color scanning electron microscope (SEM) images of the end facet (top) of the bus waveguide with a slope angle of $60^{\circ}$ and the smooth sidewall in the coupling region (bottom) of the fabricated LN microring. 
\subsection{Quasi-phase matching design and periodic poling implementation}
\begin{figure}[htbp]
\centering
\includegraphics[width=\linewidth]{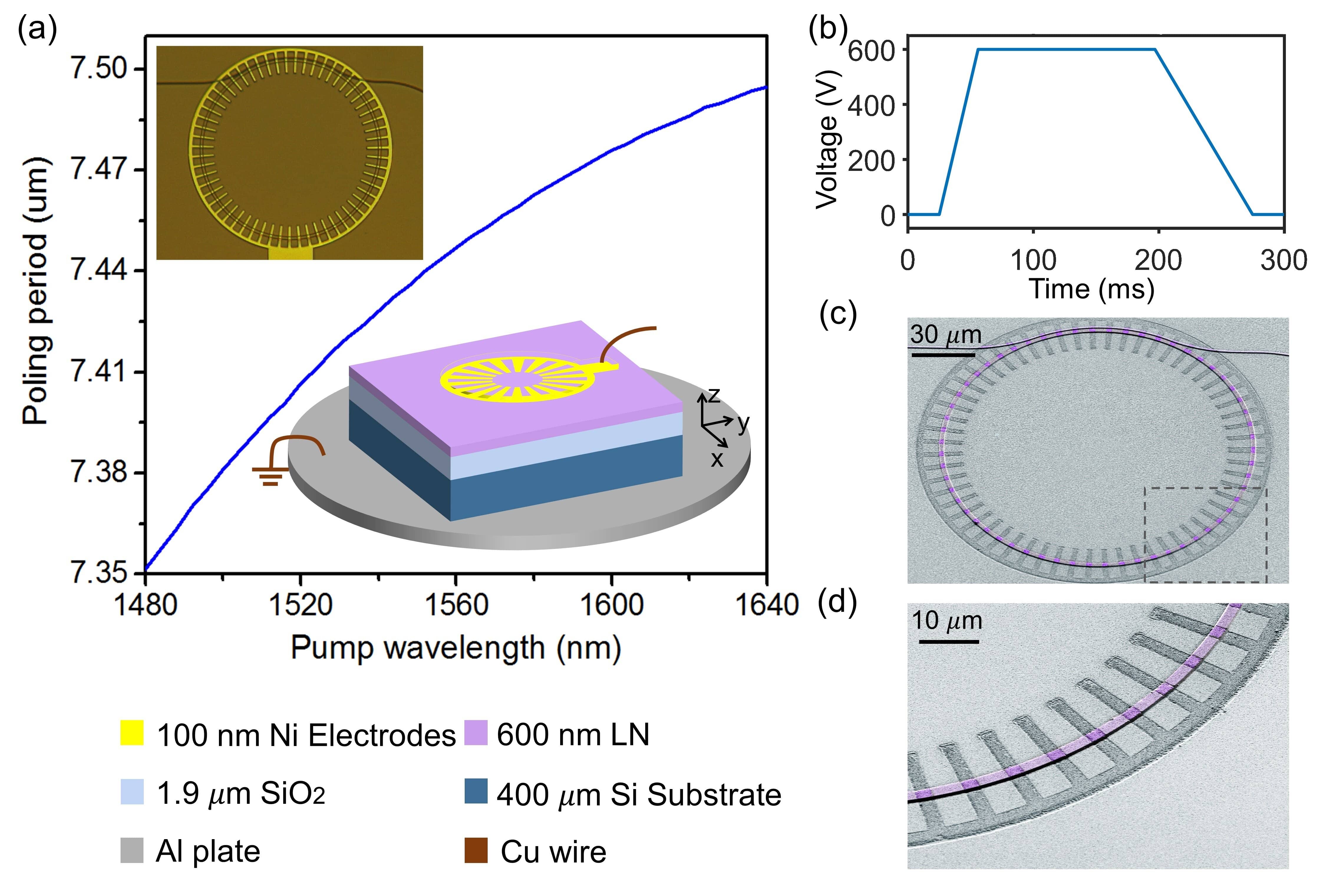}
\caption{(a) Numerical simulation of the poling period for QPM between the pump TE$_{00}$ and SH TM$_{00}$ modes using the Sellmeier equation for congruent LN in \cite{Zelmon:97}. The top inset shows a microscope image of the radial poling electrodes deposited on top of the LN microring. The bottom inset depicts the schematic diagram of the poling setup for z-cut LNOI device. (b) Applied poling pulse shape. (c-d) False-color SEM images of a PPLN microring resonator etched with hydrofluoric acid and its zoomed view, revealing a poling duty cycle of $\sim$35\%. Dark purple: inverted domain with downward z axis. Light purple: uninverted domain with upward z axis.}
\label{fig:poling}
\end{figure} 
To maximize the nonlinear coupling strength $g$ in Eq.\,(\ref{eq2}), a first-order QPM is implemented by periodically poling the fabricated LN microring along its radial orientation. Combining the QPM ($m_{b}-2m_{a}-M=0$) and dual-resonance ($\omega_{p}$\,=\,$\omega_{a}$\,=\,$\frac{\omega_{b}}{2}$, or equivalently $\lambda_{p}$\,=\,$\lambda_{a}$\,=\,2$\lambda_{b}$) conditions, the required poling period is derived as $\Lambda$\,=\,$\frac{\lambda_{a}}{2\cdot(n_{\mathrm{ring},b}-n_{\mathrm{ring},a})}$. Figure \ref{fig:poling}(a) plots the calculated $\Lambda$ for the cavity-enhanced TE$_{00}$-to-TM$_{00}$ SHG conversion, where a $\Lambda$ of approximately 7.44\,$\mu$m is obtained at a pump wavelength of 1.55\,$\mu$m. For comparison, $\Lambda$ for the TM$_{00}$-to-TM$_{00}$ conversion is estimated to be $\sim$2.87\,$\mu$m. Owing to the experimental feasibility of a large poling period as well as the potential high Q-factor of the pump TE$_{00}$ mode \cite{PhysRevApplied.11.034026,he2018self}, our work strategically focuses on the TE$_{00}$-to-TM$_{00}$ SHG conversion utilizing the $d_{31}$ coefficient ($\sim$3.2 pm/V) instead of the TM$_{00}$-to-TM$_{00}$ conversion utilizing $d_{33}$ ($\sim$19.5 pm/V) \cite{Shoji:97}.

To start the poling process, the radial nickel electrodes (top inset of Fig.\,\ref{fig:poling}(a)) were initially patterned on top of the LN microring using electron-beam lithography and lift-off processes. The poling setup for z-cut LN microring is schematically depicted in the bottom inset. The periodic crystal domain inversion was enabled by keeping the bottom aluminum plate as the electrical ground while applying six high-voltage pulses as shown in Fig.\,\ref{fig:poling}(b), on the electrodes at an elevated temperature of 250\,$^\circ\mathrm{C}$ \cite{polinghighT}. To characterize the domain inversion, the PPLN microring was etched in hydrofluoric acid \cite{B106279B} after removing the nickel electrodes. Figures\,\ref{fig:poling}(c-d) show false-color SEM images of the fabricated PPLN microring, revealing a periodic domain structure with the dark purple regions corresponding to the inverted domains. A poling period of 7.46\,$\mu$m and a duty cycle of $\sim$35\% are also extracted from the selectively etched pattern. More details of the poling process of bare z-cut LNOI wafer and domain characterization are presented in Supplement 1, Section 1.

\section{Results and discussion}

\begin{figure}[htbp]
\centering
\includegraphics[width=\linewidth]{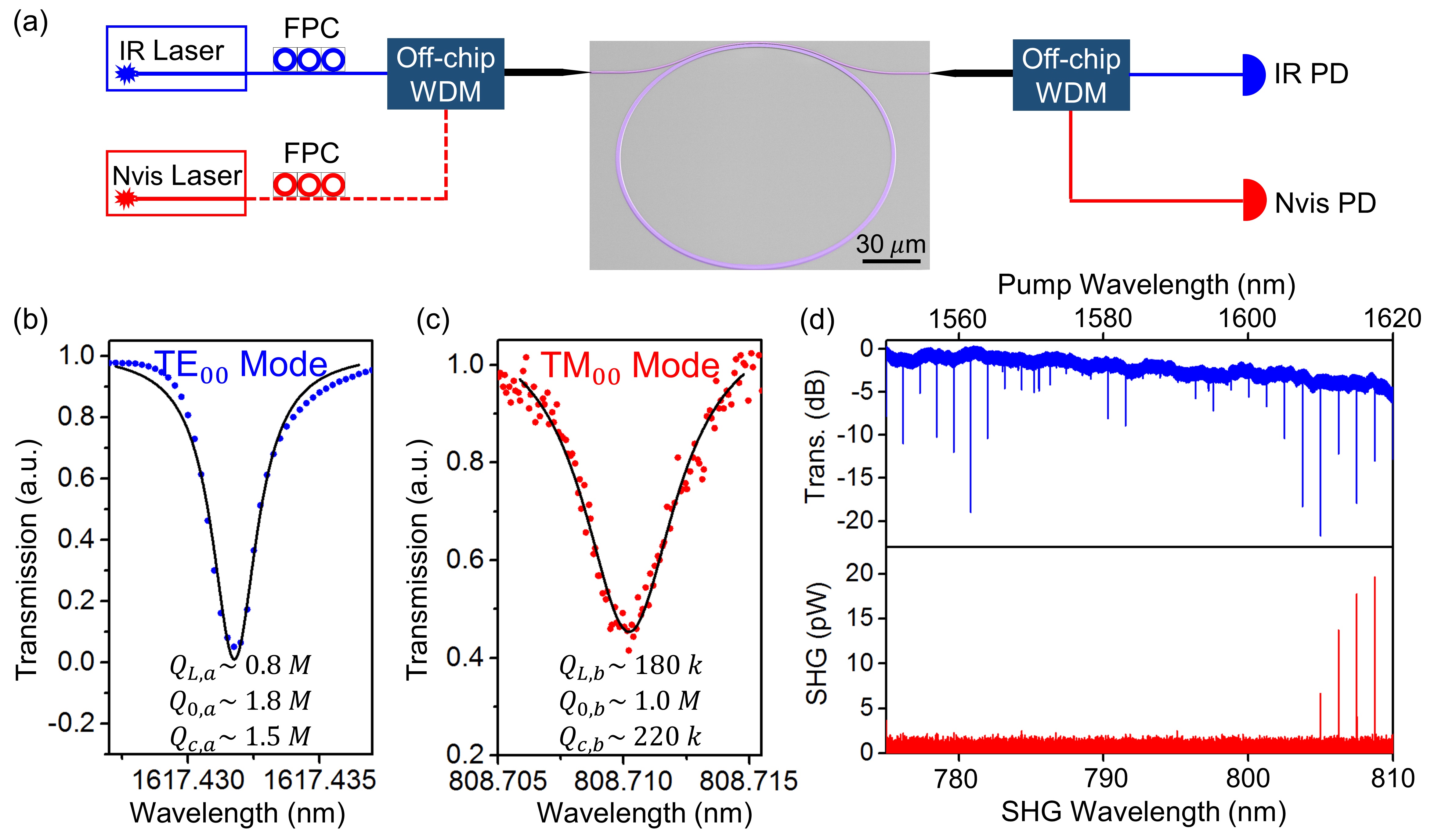}
\caption{(a) Illustration of the experimental setup. The telecom and near-visible light sources were selectively turned on for optical Q measurements, while only the telecom laser was on during the SHG measurement. (b-c) Resonance spectra of telecom TE$_{00}$ (left) and near-visible TM$_{00}$ (right) modes with extracted loaded ($Q_{L}$), intrinsic ($Q_{0}$) and coupling ($Q_{c}$) Q values (see Supplement 1, Section 2 for details). (d) Quasi-phase-matched SHG signals near 805–810\,nm (bottom panel) when sweeping a TE-polarized pump laser (blue) across a number of telecom resonances (top panel).}
\label{fig:setup}
\end{figure}
Figure\,\ref{fig:setup}(a) illustrates the experimental setup for characterizing the performance of the PPLN microring. Tunable telecom (Santec TSL710) and near-visible laser (M2 SolsTiS, 700–1000\,nm) sources were combined by a wavelength division multiplexer (WDM) and guided onto the chip by a lensed fiber. Two fiber polarization controllers (FPCs) were utilized to select horizontally and vertically polarized telecom and near-visible incident lights, respectively. The telecom and near-visible outputs were separated by a second WDM and measured by their respective photodetectors (PDs). The transmission spectra of the telecom TE$_{00}$ and near-visible TM$_{00}$ modes for optimized SHG are respectively depicted in Figs.\,\ref{fig:setup}(b) and \ref{fig:setup}(c), indicating that the pump mode is nearly critically coupled with a loaded Q-factor ($Q_{L,a}$) of 8.0\,$\times\,10^5$, while the SH TM$_{00}$ mode is over-coupled (see Supplement 1, section 2) with a $Q_{L,b}$ of 1.8\,$\times\,10^5$. The corresponding intrinsic and coupling Q-factors are extracted and displayed in the insets. Based on design parameters, their azimuthal mode numbers were respectively simulated to be $m_{a}=540$ and $m_b=1139$, which satisfy the QPM condition ($m_{b}-2 m_{a}-M=0$) when accounting for a radial poling structure with $M=\frac{2\pi R}{\Lambda}=59$. Figure\,\ref{fig:setup}(d) shows typical SHG spectra obtained by sweeping the TE-polarized pump across a number of resonances with an external input power of -13\,dBm. The distinct SHG observed was only attributed to the QPM process since our microring geometry was specifically designed to avoid possible MPM processes and no near-visible output signals were detected before the poling process (see Supplement 1, Section 3). 

\begin{figure}[htbp]
\centering
\includegraphics[width=\linewidth]{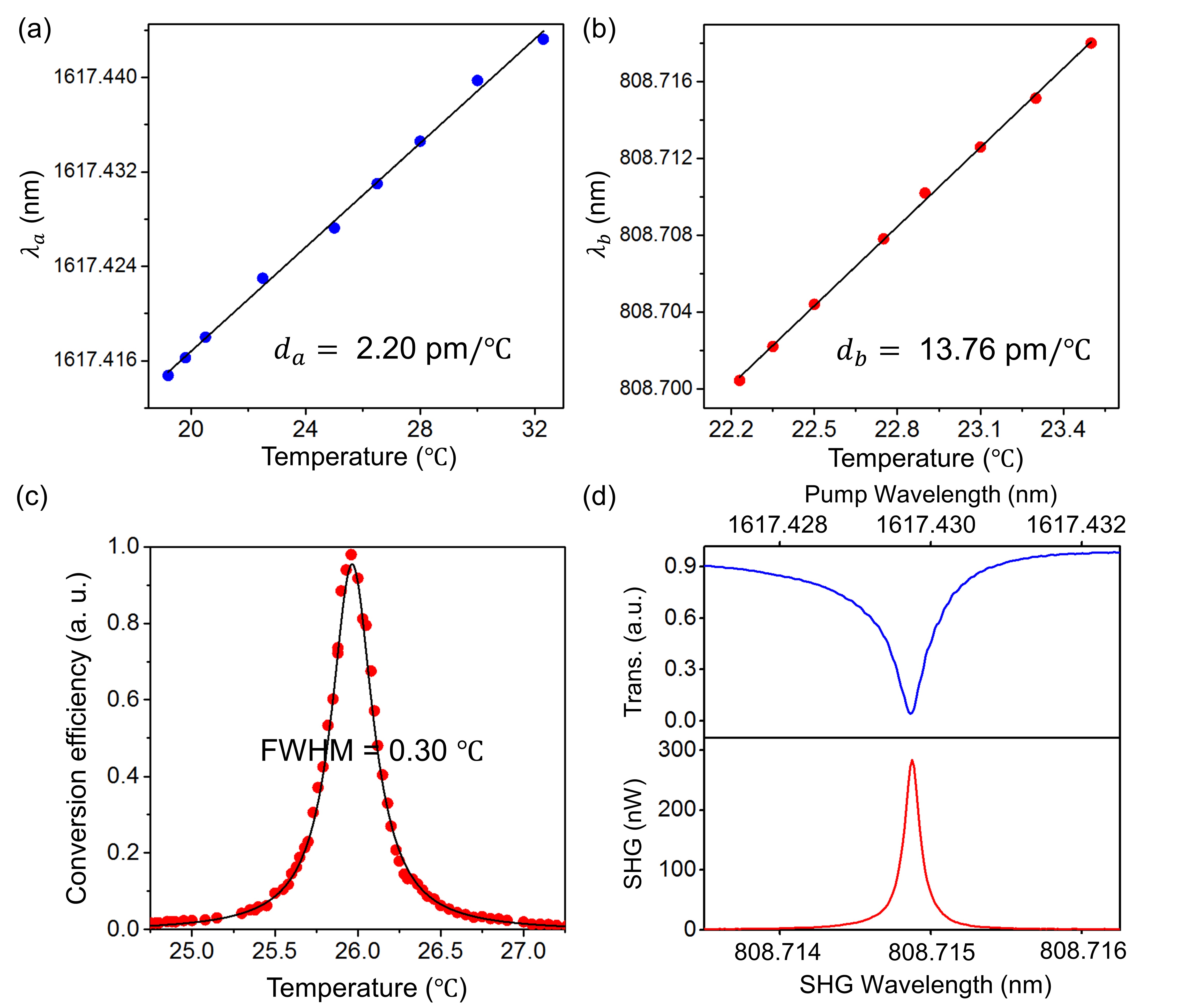}
\caption{(a-b) Measured resonance shifts versus the temperature for the pump and SH modes, respectively. (c) Recorded SHG conversion efficiency when tuning the temperature across one resonance. A Lorentzian fit is used to extract a temperature FWHM of $0.30\,^\circ\mathrm{C}$. (d) Zoom-in spectra of the pump resonance and corresponding SH response at the optimal temperature of (c).}
\label{fig:SHG}
\end{figure}
Since LN microrings exhibit both material and geometric dispersion, typically there exists a frequency mismatch between the telecom and corresponding SH near-visible cavity modes. The distinct thermal shifts of the interacting two modes then allow us to leverage temperature tuning using an external heater to perfectly fulfill the dual-resonance condition, thus further optimizing the SHG output. As shown in Figs.\,\ref{fig:SHG}(a-b), both of the telecom and near-visible resonances shift to longer wavelengths with the increasing temperature due to the thermo-optic effect and thermal expansion \cite{Carmon:04}. Moreover, the measured thermal shifts are linearly fitted by $\lambda_{a}(T) = \lambda_{{a,0}} + d_{a} \cdot T$, and $\lambda_{b}(T) = \lambda_{{b,0}} + d_{b} \cdot T$ with respective slopes of $d_{a}=2.2\, \mathrm{pm}/^\circ\mathrm{C}$ and $d_{b}=13.8\, \mathrm{pm}/^\circ\mathrm{C}$. The much larger slope for the near-visible TM$_{00}$ mode is due to the significant thermo-optic birefringence of LN \cite{doi:10.1063/1.1988987}. Considering the temperature dependence of the cavity modes and assuming the on-resonance pump wavelength, the SHG efficiency as a function of the temperature can be described as an expansion of Eq.\,(\ref{eq1})
\begin{equation}
\eta \approx \frac{16 g^{2} \frac{\kappa_{c, b}}{\hbar \omega_{a}}\left(\frac{2 \kappa_{c, a}}{\kappa_{a}^{2}}\right)^{2}}{\left[\frac{2 \pi c}{\lambda_{a} \lambda_{b}}\left(\Delta \lambda_0+\Delta d \cdot T\right)\right]^{2}+\kappa_{b}^{2}},
\end{equation}
\label{eq7}
where $\Delta \lambda_0=\lambda_{a,0}-2 \lambda_{b,0}$ and $\Delta d=d_{a}-2 d_{b}$. The temperature dependence of $\eta$ evidently exhibits a Lorentzian shape with a full width at half maximum (FWHM) determined by $\Delta T=\frac{\kappa_{b} \lambda_{a} \lambda_{b}}{\pi c \,\Delta d}=\frac{\lambda_{a}}{Q_{L,b}\, \Delta d}$. Substituting the experimentally measured $\lambda_{a}$, $Q_{L,b}$ and $\Delta d$, we calculated a $\Delta T= 0.35\,^\circ\mathrm{C}$, which agrees well with the experimental $\Delta T=0.30\,^\circ\mathrm{C}$ in Fig.\,\ref{fig:SHG}(c). Figure\,\ref{fig:SHG}(d) highlights the measured pump transmission and the corresponding SH response at the optimal temperature depicted by Fig.\,\ref{fig:SHG}(c), showing a maximum external SHG power of 280\,nW from the PPLN microring with an external pump power of 360\,$\mu$W. 

\begin{figure}[htbp]
\centering
\includegraphics[width=\linewidth]{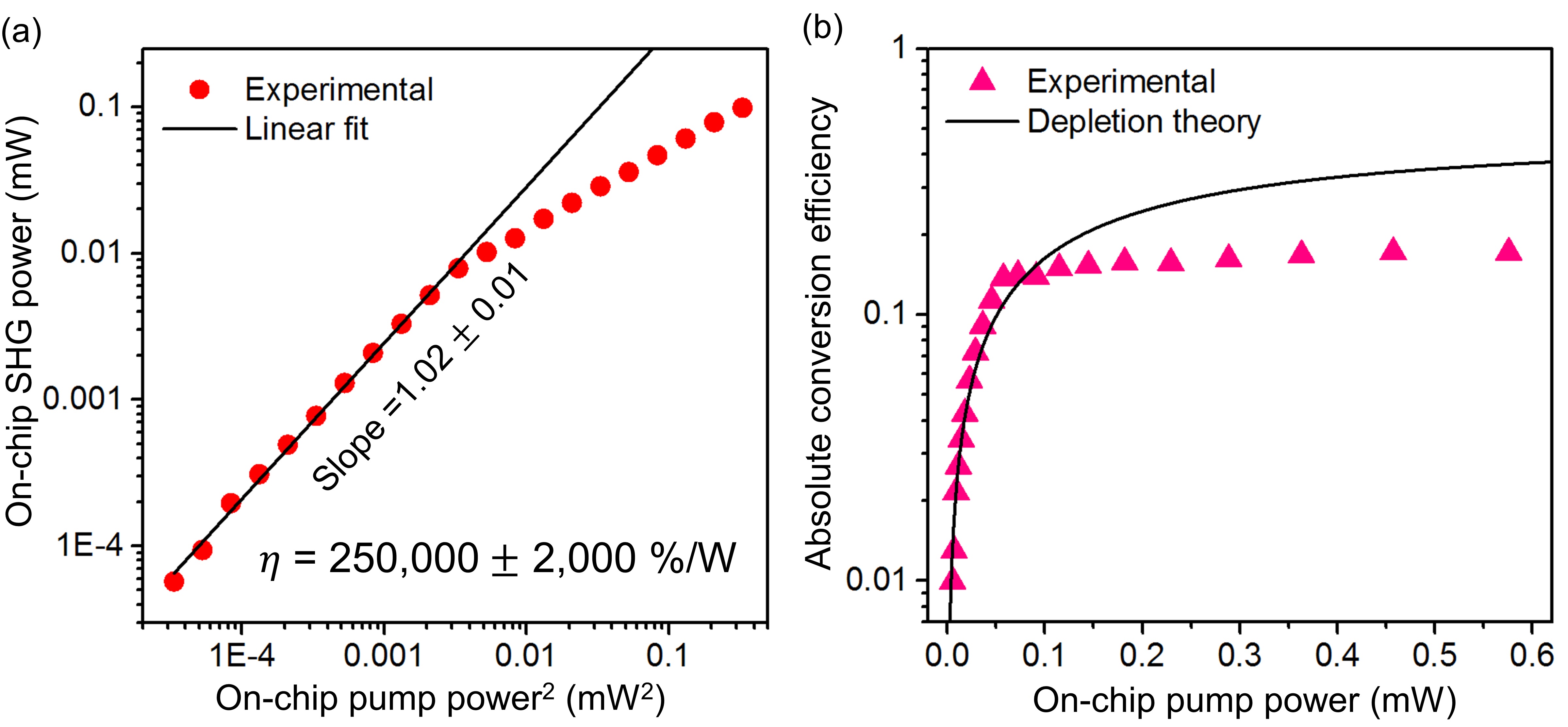}
\caption{(a) $P_{SHG}$-$P^2_{p}$ relation. A linear fit is applied to the experimental data in the low power region. A fitted slope of 1.02 implies a quadratic dependence of SHG power on the pump power. A SHG conversion efficiency of 250,000\,\%/W is extracted. (b) Absolute conversion efficiency as a function of pump power, including the experimental data and theoretical fit \cite{threewavemixing,PhysRevLett.92.043903}.}
\label{fig:efficiency}
\end{figure}

Finally, the power dependence of SHG was investigated. As shown in Fig.\,\ref{fig:efficiency}(a), in the undepleted pump regime ($P_{p}\,<\,100\,\mu$W), a linear-fitted slope of 1.02 matches the theoretical prediction based on Eq.\,(\ref{eq1}) and an on-chip SHG efficiency $\eta$ of 250,000\,\%/W is further derived. The calibration method of the on-chip pump, SHG powers and conversion efficiency $\eta$ is detailed in Supplement 1, Section 4. With an extracted $\chi_\mathrm{eff}^{(2)}$ of 3.75 pm/V (Eq.\,(\ref{eq3})) and a numerically simulated $\zeta$ of 0.74\,$/\mu$m (Eq.\,(\ref{eq4})), the theoretical $g$ factor of our device is calculated to be 1.46 MHz using Eq.\,(\ref{eq2}). By incorporating the experimentally measured Q-factors, the theoretically achievable SHG efficiency in our PPLN microring device in the non-pump depleted regime is 2,220,000\,\%/W as predicted by Eq.\,(\ref{eq1}). The one-order of magnitude discrepancy between the experimental result and theoretical prediction is possibly due to nonuniformity inherent to nanofabrication at different azimuthal angle of the microring \cite{PhysRevApplied.11.034026} as well as imperfect domain poling quality, i.e. inhomogeneity of the inverted domain, and deviation of domain boundaries (see Supplement 1, Section 1).

Figure\,\ref{fig:efficiency}(b) presents the pump power dependence of the absolute conversion efficiency, showing a measured saturation value of 15\% when the on-chip pump power reaches 115 $\mu$W. Based on the theoretical modelling considering the pump depletion \cite{threewavemixing,PhysRevLett.92.043903}, the maximum absolute conversion efficiency is $\left(\frac{P_{SHG}}{P_p}\right)_\mathrm{max}=\frac{Q_{L,a}}{Q_{c,a}}\cdot \frac{Q_{L,b}}{Q_{c,b}}$ at a saturation pump power of $P_{p}^\mathrm{{max}}=\frac{16}{\eta}\cdot \frac{Q_{L,a}}{Q_{c,a}}\cdot \frac{Q_{L,b}}{Q_{c,b}}$. On condition that both pump and SH modes are strongly over-coupled ($Q_{0,a(b)} \gg Q_{c,a(b)}$), the optimal absolute conversion efficiency approaches 100\% at the cost of increased saturation power. In our case, according to the experimentally extracted $\frac{Q_{L,a}}{Q_{c,a}}=0.53$, $\frac{Q_{L,b}}{Q_{c,b}}=0.82$, and $\eta$ of 250,000\,\%/W, the absolute conversion efficiency of the presented device is predicted to reach 44\% at a saturation power of 2.8\,mW. Nevertheless, our experimental observation (pink triangles) indicates a much lower saturation power than the theoretical prediction (dark line). Such discrepancy was also observed in previous studies of SHG in whispering gallery mode LN resonators \cite{PhysRevLett.92.043903,PhysRevLett.104.153901} and may be attributed to the enhanced photorefractive (PR) effect with the high intracavity power. As the pump power increases, the PR effect induces wavelength-dependent blueshifts of the cavity resonances \cite{he2018self,Jiang:17}, thereby creating a frequency mismatch of the pump and SH resonances, which hinders the maximum obtainable conversion efficiency at a preselected optimal temperature. Further studies will be required to quantify and compensate the PR effect via the dynamical thermal or electrical tuning in order to improve the device performance in the high power regime.

\section{Conclusion}
We demonstrate an efficient and compact $\chi^{(2)}$ frequency doubler from the telecom to near-visible band via quasi-phase matching in dual-resonant, periodically poled LN microring resonators. An on-chip SHG conversion efficiency of up to 250,000\,\%/W is recorded in the undepleted pump regime, which is a state-of-the-art value across all current integrated photonics platforms. The absolute conversion efficiency reaches 15\% at a low pump power of 115\,$\mu$W. Such high-efficiency SHG and flexibility in defining the domain switching patterns in z-cut PPLN microrings suggest its great promise for cavity-enhanced nonlinear $\chi^{(2)}$ processes. Further optimizations of the poling process, optical Q-factors and coupling conditions as well as suppression of the PR effect for a device that takes full advantage of the $d_{33}$ based on the z-cut PPLN microring platform would be a focus of our future work, and thereby an ultimate-efficiency SHG conversion could be envisioned.
 
\textit{Note added:} During the peer review of this paper, a memorandum of high-efficiency second harmonic generation based on periodically poled x-cut LN nanophotonic waveguide within a racetrack resonator was reported in \cite{Chen:19}.

\section*{Funding Information}
This work is supported by Department of Energy (DE-SC0019406), with partial support from National Science Foundation (EFMA-1640959), Army Research Office (W911NF-18-1-0020), Defense Advanced Research Projects Agency (DARPA) (HR0011-16-C-0118), David and Lucile Packard Foundation (2009-34719).

\section*{acknowledgment}
The facilities used for device fabrication were supported by the Yale SEAS cleanroom and Yale Institute for Nanoscience and Quantum Engineering. The authors thank Dr. Michael Rooks, Dr. Yong Sun, Sean Rinehart, and Kelly Woods for assistance in device fabrication.\\

\noindent{See Supplement 1 for supporting content.}




\bibliography{references}

\begin{thebibliography}{10}
\newcommand{\enquote}[1]{``#1''}

\bibitem{RevModPhys.75.325}
S.~T. Cundiff and J.~Ye, \enquote{Colloquium: Femtosecond optical frequency
  combs,} {\protect\JournalTitle{Rev. Mod. Phys.}} \textbf{75}, 325--342
  (2003).

\bibitem{synthesizer}
D.~T. Spencer, T.~Drake, T.~C. Briles, J.~Stone, L.~C. Sinclair, C.~Fredrick,
  Q.~Li, D.~Westly, B.~R. Ilic, A.~Bluestone, N.~Volet, T.~Komljenovic,
  L.~Chang, S.~H. Lee, D.~Y. Oh, M.-G. Suh, K.~Y. Yang, M.~H.~P. Pfeiffer,
  T.~J. Kippenberg, E.~Norberg, L.~Theogarajan, K.~Vahala, N.~R. Newbury,
  K.~Srinivasan, J.~E. Bowers, S.~A. Diddams, and S.~B. Papp, \enquote{An
  optical-frequency synthesizer using integrated photonics,}
  {\protect\JournalTitle{Nature}} \textbf{557}, 81--85 (2018).

\bibitem{Newman:19}
Z.~L. Newman, V.~Maurice, T.~Drake, J.~R. Stone, T.~C. Briles, D.~T. Spencer,
  C.~Fredrick, Q.~Li, D.~Westly, B.~R. Ilic, B.~Shen, M.-G. Suh, K.~Y. Yang,
  C.~Johnson, D.~M.~S. Johnson, L.~Hollberg, K.~J. Vahala, K.~Srinivasan, S.~A.
  Diddams, J.~Kitching, S.~B. Papp, and M.~T. Hummon, \enquote{Architecture for
  the photonic integration of an optical atomic clock,}
  {\protect\JournalTitle{Optica}} \textbf{6}, 680--685 (2019).

\bibitem{SHG:image}
Y.~R. Shen, \enquote{Surface properties probed by second-harmonic and
  sum-frequency generation,} {\protect\JournalTitle{Nature}} \textbf{337},
  519--525 (1989).

\bibitem{Han:05}
M.~Han, G.~Giese, and J.~F. Bille, \enquote{Second harmonic generation imaging
  of collagen fibrils in cornea and sclera,} {\protect\JournalTitle{Opt.
  Express}} \textbf{13}, 5791--5797 (2005).

\bibitem{xiang:light}
X.~Guo, C.-L. Zou, C.~Schuck, H.~Jung, R.~Cheng, and H.~X. Tang,
  \enquote{Parametric down-conversion photon-pair source on a nanophotonic
  chip,} {\protect\JournalTitle{Light Science Applications}} \textbf{6}, e16249
  (2017).

\bibitem{quantuminfo}
S.~Tanzilli, W.~Tittel, M.~Halder, O.~Alibart, P.~Baldi, N.~Gisin, and
  H.~Zbinden, \enquote{A photonic quantum information interface,}
  {\protect\JournalTitle{Nature}} \textbf{437}, 116--120 (2005).

\bibitem{161322}
M.~M. {Fejer}, G.~A. {Magel}, D.~H. {Jundt}, and R.~L. {Byer},
  \enquote{Quasi-phase-matched second harmonic generation: tuning and
  tolerances,} {\protect\JournalTitle{IEEE Journal of Quantum Electronics}}
  \textbf{28}, 2631--2654 (1992).

\bibitem{Levy:11}
J.~S. Levy, M.~A. Foster, A.~L. Gaeta, and M.~Lipson, \enquote{Harmonic
  generation in silicon nitride ring resonators,} {\protect\JournalTitle{Opt.
  Express}} \textbf{19}, 11415--11421 (2011).

\bibitem{xxxue:17}
X.~Xue, F.~Leo, Y.~Xuan, J.~A. Jaramillo-Villegas, P.-H. Wang, D.~E. Leaird,
  M.~Erkintalo, M.~Qi, and A.~M. Weiner, \enquote{Second-harmonic-assisted
  four-wave mixing in chip-based microresonator frequency comb generation,}
  {\protect\JournalTitle{Light Science Applications}} \textbf{6}, e16253
  (2017).

\bibitem{doi:10.1002/lpor.201800149}
L.~Chang, A.~Boes, X.~Guo, D.~T. Spencer, M.~J. Kennedy, J.~D. Peters,
  N.~Volet, J.~Chiles, A.~Kowligy, N.~Nader, D.~D. Hickstein, E.~J. Stanton,
  S.~A. Diddams, S.~B. Papp, and J.~E. Bowers, \enquote{Heterogeneously
  integrated {GaAs} waveguides on insulator for efficient frequency
  conversion,} {\protect\JournalTitle{Laser \& Photonics Reviews}} \textbf{12},
  1800149 (2018).

\bibitem{kuo2016}
P.~S. Kuo, J.~Bravo-Abad, and G.~S. Solomon, \enquote{Second-harmonic
  generation using \={4}-quasi-phasematching in a {GaAs}
  whispering-gallery-mode microcavity,} {\protect\JournalTitle{Nature
  Communications}} \textbf{5}, 3109 (2014).

\bibitem{Guo:16}
X.~Guo, C.-L. Zou, and H.~X. Tang, \enquote{Second-harmonic generation in
  aluminum nitride microrings with 2500\%/{W} conversion efficiency,}
  {\protect\JournalTitle{Optica}} \textbf{3}, 1126--1131 (2016).

\bibitem{Surya:18}
J.~B. Surya, X.~Guo, C.-L. Zou, and H.~X. Tang, \enquote{Control of
  second-harmonic generation in doubly resonant aluminum nitride microrings to
  address a rubidium two-photon clock transition,} {\protect\JournalTitle{Opt.
  Lett.}} \textbf{43}, 2696--2699 (2018).

\bibitem{alexSHG}
A.~W. Bruch, X.~Liu, X.~Guo, J.~B. Surya, Z.~Gong, L.~Zhang, J.~Wang, J.~Yan,
  and H.~X. Tang, \enquote{17000\%/{W} second-harmonic conversion efficiency in
  single-crystalline aluminum nitride microresonators,}
  {\protect\JournalTitle{Applied Physics Letters}} \textbf{113}, 131102 (2018).

\bibitem{chang2016thin}
L.~Chang, Y.~Li, N.~Volet, L.~Wang, J.~Peters, and J.~E. Bowers, \enquote{Thin
  film wavelength converters for photonic integrated circuits,}
  {\protect\JournalTitle{Optica}} \textbf{3}, 531--535 (2016).

\bibitem{Wolf:18}
R.~Wolf, Y.~Jia, S.~Bonaus, C.~S. Werner, S.~J. Herr, I.~Breunig, K.~Buse, and
  H.~Zappe, \enquote{Quasi-phase-matched nonlinear optical frequency conversion
  in on-chip whispering galleries,} {\protect\JournalTitle{Optica}} \textbf{5},
  872--875 (2018).

\bibitem{PhysRevApplied.11.034026}
R.~Luo, Y.~He, H.~Liang, M.~Li, J.~Ling, and Q.~Lin, \enquote{Optical
  parametric generation in a lithium niobate microring with modal phase
  matching,} {\protect\JournalTitle{Phys. Rev. Applied}} \textbf{11}, 034026
  (2019).

\bibitem{Luo:18}
R.~Luo, Y.~He, H.~Liang, M.~Li, and Q.~Lin, \enquote{Highly tunable efficient
  second-harmonic generation in a lithium niobate nanophotonic waveguide,}
  {\protect\JournalTitle{Optica}} \textbf{5}, 1006--1011 (2018).

\bibitem{Wang:18}
C.~Wang, C.~Langrock, A.~Marandi, M.~Jankowski, M.~Zhang, B.~Desiatov, M.~M.
  Fejer, and M.~Lon\v{c}ar, \enquote{Ultrahigh-efficiency wavelength conversion
  in nanophotonic periodically poled lithium niobate waveguides,}
  {\protect\JournalTitle{Optica}} \textbf{5}, 1438--1441 (2018).

\bibitem{doi:10.1002/lpor.201800288}
R.~Luo, Y.~He, H.~Liang, M.~Li, and Q.~Lin, \enquote{Semi-nonlinear
  nanophotonic waveguides for highly efficient second-harmonic generation,}
  {\protect\JournalTitle{Laser \& Photonics Reviews}} \textbf{13}, 1800288
  (2019).

\bibitem{Xiong:11}
C.~Xiong, W.~Pernice, K.~K. Ryu, C.~Schuck, K.~Y. Fong, T.~Palacios, and H.~X.
  Tang, \enquote{Integrated {GaN} photonic circuits on silicon (100) for second
  harmonic generation,} {\protect\JournalTitle{Opt. Express}} \textbf{19},
  10462--10470 (2011).

\bibitem{Roland:16}
I.~Roland, M.~Gromovyi, Y.~Zeng, M.~El~Kurdi, S.~Sauvage, C.~Brimont,
  T.~Guillet, B.~Gayral, F.~Semond, J.~Y. Duboz, M.~de~Micheli, X.~Checoury,
  and P.~Boucaud, \enquote{Phase-matched second harmonic generation with
  on-chip {GaN-on-Si} microdisks,} {\protect\JournalTitle{Scientific Reports}}
  \textbf{6}, 34191 (2016).

\bibitem{Scaccabarozzi:06}
L.~Scaccabarozzi, M.~M. Fejer, Y.~Huo, S.~Fan, X.~Yu, and J.~S. Harris,
  \enquote{Enhanced second-harmonic generation in {AlGaAs/AlxOy} tightly
  confining waveguides and resonant cavities,} {\protect\JournalTitle{Opt.
  Lett.}} \textbf{31}, 3626--3628 (2006).

\bibitem{Rivoire:09}
K.~Rivoire, Z.~Lin, F.~Hatami, W.~T. Masselink, and J.~Vu\v{c}kovi\'{c},
  \enquote{Second harmonic generation in gallium phosphide photonic crystal
  nanocavities with ultralow continuous wave pump power,}
  {\protect\JournalTitle{Opt. Express}} \textbf{17}, 22609--22615 (2009).

\bibitem{Yamada:14}
S.~Yamada, B.-S. Song, S.~Jeon, J.~Upham, Y.~Tanaka, T.~Asano, and S.~Noda,
  \enquote{Second-harmonic generation in a silicon-carbide-based photonic
  crystal nanocavity,} {\protect\JournalTitle{Opt. Lett.}} \textbf{39},
  1768--1771 (2014).

\bibitem{Zhang17}
M.~Zhang, C.~Wang, R.~Cheng, A.~Shams-Ansari, and M.~Lon\v{c}ar,
  \enquote{Monolithic {ultra-high-Q} lithium niobate microring resonator,}
  {\protect\JournalTitle{Optica}} \textbf{4}, 1536--1537 (2017).

\bibitem{wang18}
C.~Wang, M.~Zhang, X.~Chen, M.~Bertrand, A.~Shams-Ansari, S.~Chandrasekhar,
  P.~Winzer, and M.~Lon{\v c}ar, \enquote{Integrated lithium niobate
  electro-optic modulators operating at {CMOS}-compatible voltages,}
  {\protect\JournalTitle{Nature}} \textbf{562}, 101--104 (2018).

\bibitem{zhang19}
M.~Zhang, C.~Wang, Y.~Hu, A.~Shams-Ansari, T.~Ren, S.~Fan, and M.~Lon{\v c}ar,
  \enquote{Electronically programmable photonic molecule,}
  {\protect\JournalTitle{Nature Photonics}} \textbf{13}, 36--40 (2019).

\bibitem{wang19}
C.~Wang, M.~Zhang, M.~Yu, R.~Zhu, H.~Hu, and M.~Loncar, \enquote{Monolithic
  lithium niobate photonic circuits for kerr frequency comb generation and
  modulation,} {\protect\JournalTitle{Nature Communications}} \textbf{10}, 978
  (2019).

\bibitem{eocomb}
M.~Zhang, B.~Buscaino, C.~Wang, A.~Shams-Ansari, C.~Reimer, R.~Zhu, J.~M. Kahn,
  and M.~Lon{\v c}ar, \enquote{Broadband electro-optic frequency comb
  generation in a lithium niobate microring resonator,}
  {\protect\JournalTitle{Nature}} \textbf{568}, 373--377 (2019).

\bibitem{he2018self}
Y.~He, Q.-F. Yang, J.~Ling, R.~Luo, H.~Liang, M.~Li, B.~Shen, H.~Wang,
  K.~Vahala, and Q.~Lin, \enquote{A self-starting bi-chromatic {LiNbO$_{3}$}
  soliton microcomb,} {\protect\JournalTitle{arXiv:1812.09610}}  (2018).

\bibitem{Desiatov:19}
B.~Desiatov, A.~Shams-Ansari, M.~Zhang, C.~Wang, and M.~Lon\v{c}ar,
  \enquote{Ultra-low-loss integrated visible photonics using thin-film lithium
  niobate,} {\protect\JournalTitle{Optica}} \textbf{6}, 380--384 (2019).

\bibitem{Luo:17}
R.~Luo, H.~Jiang, S.~Rogers, H.~Liang, Y.~He, and Q.~Lin, \enquote{On-chip
  second-harmonic generation and broadband parametric down-conversion in a
  lithium niobate microresonator,} {\protect\JournalTitle{Opt. Express}}
  \textbf{25}, 24531--24539 (2017).

\bibitem{Yu:19}
M.~Yu, B.~Desiatov, Y.~Okawachi, A.~L. Gaeta, and M.~Lon\v{c}ar,
  \enquote{Coherent two-octave-spanning supercontinuum generation in
  lithium-niobate waveguides,} {\protect\JournalTitle{Opt. Lett.}} \textbf{44},
  1222--1225 (2019).

\bibitem{Lu:19}
J.~Lu, J.~B. Surya, X.~Liu, Y.~Xu, and H.~X. Tang, \enquote{Octave-spanning
  supercontinuum generation in nanoscale lithium niobate waveguides,}
  {\protect\JournalTitle{Opt. Lett.}} \textbf{44}, 1492--1495 (2019).

\bibitem{vahalamicrocavity}
K.~J. Vahala, \enquote{Optical microcavities,} {\protect\JournalTitle{Nature}}
  \textbf{424}, 839--846 (2003).

\bibitem{Wang:14}
C.~Wang, M.~J. Burek, Z.~Lin, H.~A. Atikian, V.~Venkataraman, I.-C. Huang,
  P.~Stark, and M.~Lon\v{c}ar, \enquote{Integrated high quality factor lithium
  niobate microdisk resonators,} {\protect\JournalTitle{Opt. Express}}
  \textbf{22}, 30924--30933 (2014).

\bibitem{Chen:18}
J.-Y. Chen, Y.~M. Sua, H.~Fan, and Y.-P. Huang, \enquote{Modal phase matched
  lithium niobate nanocircuits for integrated nonlinear photonics,}
  {\protect\JournalTitle{OSA Continuum}} \textbf{1}, 229--242 (2018).

\bibitem{PhysRevApplied.6.014002}
J.~Lin, Y.~Xu, J.~Ni, M.~Wang, Z.~Fang, L.~Qiao, W.~Fang, and Y.~Cheng,
  \enquote{Phase-matched second-harmonic generation in an on-chip
  {${\mathrm{L}\mathrm{i}\mathrm{NbO}}_{3}$} microresonator,}
  {\protect\JournalTitle{Phys. Rev. Applied}} \textbf{6}, 014002 (2016).

\bibitem{PhysRevLett.122.173903}
J.~Lin, N.~Yao, Z.~Hao, J.~Zhang, W.~Mao, M.~Wang, W.~Chu, R.~Wu, Z.~Fang,
  L.~Qiao, W.~Fang, F.~Bo, and Y.~Cheng, \enquote{Broadband quasi-phase-matched
  harmonic generation in an on-chip monocrystalline lithium niobate microdisk
  resonator,} {\protect\JournalTitle{Phys. Rev. Lett.}} \textbf{122}, 173903
  (2019).

\bibitem{HUM2007180}
D.~S. Hum and M.~M. Fejer, \enquote{Quasi-phasematching,}
  {\protect\JournalTitle{Comptes Rendus Physique}} \textbf{8}, 180 -- 198
  (2007).

\bibitem{Hosseini:10}
E.~S. Hosseini, S.~Yegnanarayanan, A.~H. Atabaki, M.~Soltani, and A.~Adibi,
  \enquote{Systematic design and fabrication of {high-Q} single-mode
  pulley-coupled planar silicon nitride microdisk resonators at visible
  wavelengths,} {\protect\JournalTitle{Opt. Express}} \textbf{18}, 2127--2136
  (2010).

\bibitem{Liu:17}
X.~Liu, C.~Sun, B.~Xiong, L.~Wang, J.~Wang, Y.~Han, Z.~Hao, H.~Li, Y.~Luo,
  J.~Yan, T.~Wei, Y.~Zhang, and J.~Wang, \enquote{Aluminum nitride-on-sapphire
  platform for integrated {high-Q} microresonators,}
  {\protect\JournalTitle{Opt. Express}} \textbf{25}, 587--594 (2017).

\bibitem{Zelmon:97}
D.~E. Zelmon, D.~L. Small, and D.~Jundt, \enquote{Infrared corrected sellmeier
  coefficients for congruently grown lithium niobate and 5 mol.\% magnesium
  oxide--doped lithium niobate,} {\protect\JournalTitle{J. Opt. Soc. Am. B}}
  \textbf{14}, 3319--3322 (1997).

\bibitem{Shoji:97}
I.~Shoji, T.~Kondo, A.~Kitamoto, M.~Shirane, and R.~Ito, \enquote{Absolute
  scale of second-order nonlinear-optical coefficients,}
  {\protect\JournalTitle{J. Opt. Soc. Am. B}} \textbf{14}, 2268--2294 (1997).

\bibitem{polinghighT}
H.~Ishizuki, I.~Shoji, and T.~Taira, \enquote{Periodical poling characteristics
  of congruent {MgO:LiNbO$_3$} crystals at elevated temperature,}
  {\protect\JournalTitle{Applied Physics Letters}} \textbf{82}, 4062--4064
  (2003).

\bibitem{B106279B}
C.~L. Sones, S.~Mailis, W.~S. Brocklesby, R.~W. Eason, and J.~R. Owen,
  \enquote{Differential etch rates in z-cut {LiNbO$_3$} for variable
  {HF/HNO$_3$} concentrations,} {\protect\JournalTitle{J. Mater. Chem.}}
  \textbf{12}, 295--298 (2002).

\bibitem{Carmon:04}
T.~Carmon, L.~Yang, and K.~J. Vahala, \enquote{Dynamical thermal behavior and
  thermal self-stability of microcavities,} {\protect\JournalTitle{Opt.
  Express}} \textbf{12}, 4742--4750 (2004).

\bibitem{doi:10.1063/1.1988987}
L.~Moretti, M.~Iodice, F.~G. Della~Corte, and I.~Rendina, \enquote{Temperature
  dependence of the thermo-optic coefficient of lithium niobate, from 300 to
  515 {K} in the visible and infrared regions,} {\protect\JournalTitle{Journal
  of Applied Physics}} \textbf{98}, 036101 (2005).

\bibitem{threewavemixing}
I.~Breunig, \enquote{Three-wave mixing in whispering gallery resonators,}
  {\protect\JournalTitle{Laser \& Photonics Reviews}} \textbf{10}, 569--587
  (2016).

\bibitem{PhysRevLett.92.043903}
V.~S. Ilchenko, A.~A. Savchenkov, A.~B. Matsko, and L.~Maleki,
  \enquote{Nonlinear optics and crystalline whispering gallery mode cavities,}
  {\protect\JournalTitle{Phys. Rev. Lett.}} \textbf{92}, 043903 (2004).

\bibitem{PhysRevLett.104.153901}
J.~U. F\"urst, D.~V. Strekalov, D.~Elser, M.~Lassen, U.~L. Andersen,
  C.~Marquardt, and G.~Leuchs, \enquote{Naturally phase-matched second-harmonic
  generation in a whispering-gallery-mode resonator,}
  {\protect\JournalTitle{Phys. Rev. Lett.}} \textbf{104}, 153901 (2010).

\bibitem{Jiang:17}
H.~Jiang, R.~Luo, H.~Liang, X.~Chen, Y.~Chen, and Q.~Lin, \enquote{Fast
  response of photorefraction in lithium niobate microresonators,}
  {\protect\JournalTitle{Opt. Lett.}} \textbf{42}, 3267--3270 (2017).

\bibitem{Chen:19}
J.-Y. Chen, Z.-H. Ma, Y.~M. Sua, Z.~Li, C.~Tang, and Y.-P. Huang,
  \enquote{Ultra-efficient frequency conversion in quasi-phase-matched lithium
  niobate microrings,} {\protect\JournalTitle{Optica}} \textbf{6}, 1244--1245
  (2019).

\end{thebibliography}

\bibliographyfullrefs{references}


\end{document}